\documentclass[twocolumn,showpacs,preprintnumbers,amsmath,amssymb]{revtex4}
\usepackage{tipa}
\usepackage{amssymb}
\usepackage{stmaryrd}
\usepackage{txfonts}

\usepackage{graphicx}
\usepackage{dcolumn}
\usepackage{bm}

\begin{document}

\preprint{}

\title{Weak-value amplification for  Weyl-point separation in momentum space}
\author{Shizhen Chen}
\author{Chengquan Mi}
\author{Weijie Wu}
\author{Wenshuai Zhang}
\author{Hailu Luo}\email{hailuluo@hnu.edu.cn}
\author{Shuangchun Wen}
\affiliation{Laboratory for Spin Photonics, School of Physics
and Electronics, Hunan University, Changsha 410082, China}

\date{\today}

\begin{abstract}
The existence of Weyl nodes in the momentum space is a hallmark of a Weyl semimetal (WSM). A WSM can be confirmed by observing its Fermi arcs with separated Weyl nodes. In this paper, we study the spin-orbit interaction of light on the surface of WSM in the limit that the thickness is ultra-thin and the incident surface does not support Fermi arc. Our results show that the spin-dependent splitting induced by the spin-orbit interaction is related to the separation of Weyl nodes. By proposing an amplification
technique called weak measurements, the distance of the nodes can be precisely determined. This system may have application in characterizing other parameters of WSM.
\end{abstract}

\pacs{42.25.Ja, 42.25.Hz, 42.50.Xa}
\keywords{photonic spin Hall effect, weak measurements, spin-orbit
coupling}

\maketitle

\section{Introduction}\label{SecI}
Weyl fermions have been proposed and long studied in quantum field theory, but this kind of particles has not yet been observed as a fundamental particle in nature. Recent research found that Weyl fermions can appear as quasiparticles in a Weyl semimetal (WSM)~\cite{Burkov2011,Xu2011,Zyuzin2012,Xu2015,Jiang2015}. WSM is a new sate of material that hosts separated band touching points---Weyl nodes---with opposite chirality~\cite{Wan2011,Singh2012,Liu2014,Lu2015}. The Weyl nodes come in pairs in bulk Brillouin zone when time-reversal or inversion symmetry is broken. WSM has attracted much attention due to its many exotic properties induced by the Weyl nodes, such as anomalous Hall effect~\cite{Burkov2011,Yang2011}, surface states with Fermi arcs~\cite{Ojanen2013,Noh2017}, peculiar electromagnetic response~\cite{Vazifeh2013,Ukhtary2017}, and negatice magneto-resistivity~\cite{Son2013,Huang2015,Arnold2016,Zhang2017}. A WSM can be proved by observing its Fermi arcs with separated Weyl nodes. The experiment to observe Weyl nodes in TaAs or MoTe$_2$ by angle-resolved photoemission spectroscopy was recently reported~\cite{Lv2015I,Lv2015,Jiang2017}. Due to the experimental resolution and spectral linewidth, the nodes in other WSM materials, such as NbP and WTe$_2$, may become difficult to be directly observed~\cite{Belopolski2016,Bruno2016,Wu2016,Wang2016}.

\begin{figure}
\includegraphics[width=8cm]{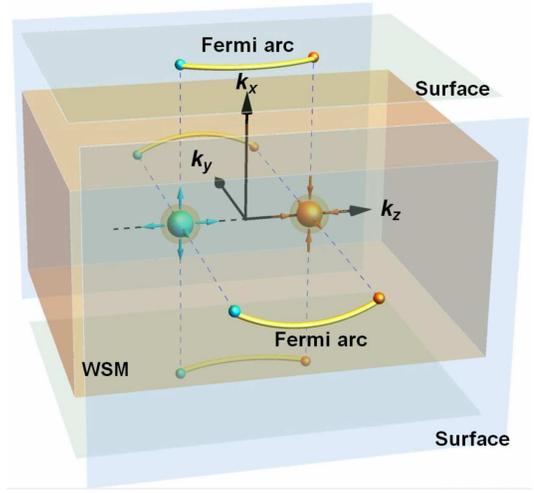}
\caption{\label{Fig1} Illustration of a Weyl semimetal with Fermi arcs on its surface of Brillouin zone connecting projections of a pair of Weyl nodes with opposite chirality. The Weyl nodes are shown as the blue and orange dots with outward and inward arrows, behaving as magnetic monopoles with topological charges. Note that the surfaces perpendicular to $k_{z}$ axis does not support arcs.}
\end{figure}

In this paper, we provide an alternative method to demonstrate the existence of Weyl-point separation in momentum space. The WSM we discuss contains only a pair of Weyl nodes with broken time reversal symmetry~\cite{Burkov2011,Kargarian2015,Ahn2017}. As illustrated in Fig.~\ref{Fig1}, the projection of the two Weyl nodes connects the ending points of Fermi arc on the Brillouin zone surface. The separation of the nodes is along the $k_{z}$ direction. We consider the electromagnetic wave incidents on the surface without Fermi arc states. The spin-orbit interaction of light on WSM occurs, which manifests itself as tiny splitting of left- and right-circular components. This phenomenon within visible wavelengths is known as photonic spin Hall effect~\cite{Onoda2004,Bliokh2006,Ling2017}. We find that the coupling in WSM is still very weak, and an amplification method called quantum weak measurements is introduced~\cite{Hosten2008,Qin2009,Luo2011,Gorodetski2012}.

\begin{figure}
\includegraphics[width=8.5cm]{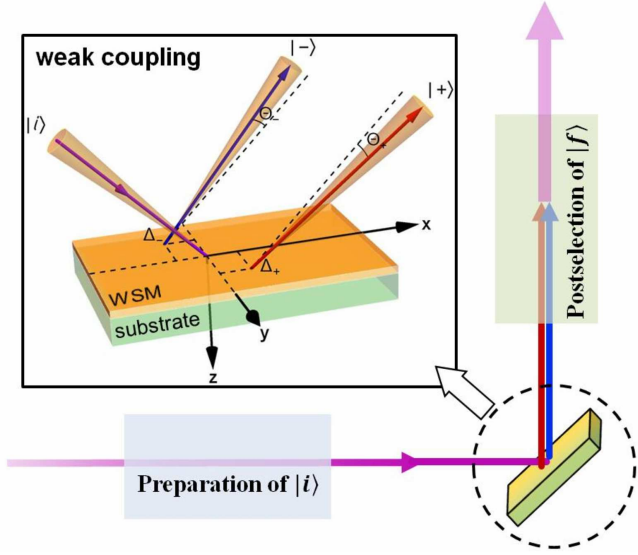}
\caption{\label{Fig2} Schematic of the weak measurement scheme to detect the weak spin-orbit coupling in a WSM-substrate system. The inset shows the splitting of left and right circularly polarized components induced by spin-orbit interaction after reflection. $\Delta$ and $\Theta$ denote the spatial and angular shifts, respectively.}
\end{figure}

The concept of weak measurements was proposed in the context of quantum mechanics~\cite{Aharonov1988,Kofman2012,Dressel2014}. There are three key elements in a weak measurement schema, namely, preselected state, postselected state, and weak coupling between the observation system and measuring pointer. The result of the whole system called weak value can be outside the eigenvalue range of the observable or even be a complex number, which is given by \begin{equation}
A_{w}=\frac{\langle{f}|\hat{A}|{i}\rangle}{\langle{f}|{i}\rangle}\label{asi},
\end{equation}
in which $\hat{A}$ is the system operator of an observable, $|{i}\rangle$ and $|{f}\rangle$ are the preselected and postselected states respectively. By making $\langle{f}|{i}\rangle\rightarrow0$, the weak value becomes very large and therefore can be utilized to amplify some tiny effect or small parameters. In our case, the polarization state of light is taken as the system state, which can be prepared by optical elements in an experiment. The spin-orbit interaction of light in the WSM-substrate system provides the weak coupling to the meter, as shown in Fig.~\ref{Fig2}. Due to the amplification effect of $A_{w}$, the outcome related to the separation of Weyl nodes makes the Fermi arcs detectable. In the following, we first discuss the spin-orbit interaction of light reflected on WSM-substrate interface.

\section{The model for spin-orbit interaction in WSM}\label{SecII}
In this section, we establish a model to describe the spin-orbit interaction in the WSM-substrate system. The WSM film with thickness $d$ is placed on the substrate. A monochromatic Gaussian beam impinging from air to WSM-substrate system is shown in the inset of Fig.~\ref{Fig2}. The optical response of the WSM changes with photon energy due to the dynamic conductivity, like the case of two dimension massive Dirac fermions~\cite{Li2013,Li2014}. Considering the low frequency limit $\omega<\omega_c$, the wavelength $\lambda$ is chosen as 633nm. $\omega_c=v_{F}k_{c}$ with $v_{F}$ and $k_{c}$ representing Fermi velocity and the momentum cutoff along ${k}_z$ axis, respectively. And the corresponding energy of the photons is 1.96ev. For the bounded beam, the polarization states of different angular spectrum components can be written as $|{{H}}({k}_i)\rangle$ and $|{{V}}({k}_i)\rangle$.
To denote central wave vector of wavepacket, the coordinate frames ($x_i,y_i,z_i$) and
($x_r,y_r,z_r$) are used, where the subscript $i$ and $ r$  respectively represent incident and reflected beam. After reflecting at the air-WSM-medium interface, the rotations of polarizations for each angular spectrum components are different. Introducing the boundary condition $k_{ry}= k_{iy}$, the total action of the reflection can be described by $[|{{H}}({k}_r)\rangle~|{{V}}({k}_r)\rangle]^T={M}_{R}[|{H}({k}_i)\rangle~|{V}({k}_i)\rangle]^T$, where ${M}_{R}$ is
\begin{eqnarray}
\left[
\begin{array}{cc}
r_{pp}-\frac{k_{ry}\cot\theta(r_{ps}-r_{sp})}{k_{0}} &r_{ps}+\frac{k_{ry}\cot\theta(r_{pp}+r_{ss})}{k_{0}} \\
r_{sp}-\frac{k_{ry}\cot\theta(r_{pp}+r_{ss})}{k_{0}} &
r_{ss}-\frac{k_{ry}\cot\theta(r_{ps}-r_{sp})}{k_{0}}
\end{array}\right]\label{MatrixRII}.
\end{eqnarray}
$r_{ab}$ is the Fresnel reflection coefficients of the WSM-substrate system with $a$ and $b$ standing for either $s$ or $p$ polarization. In order to obtain the in-plane displacement and more precisely describe the transverse splitting, we expand the Fresnel coefficients $r_{pp}$ and $r_{sp}$ to the first order in a Taylor series
expansion
\begin{eqnarray}
r_{ab}(k_{ix})&=&r_{ab}(k_{ix}=0)+k_{ix}\left[\frac{\partial
r_{ab}(k_{ix})}{\partial
k_{ix}}\right]_{k_{ix}=0}
\label{Talorkx},
\end{eqnarray}
For simplicity, we only discuss the case with horizontal incident polarization state. The state after reflection becomes
\begin{eqnarray}
 |{H}({k}_{i})\rangle&\rightarrow& \left(r_{pp}-\frac{k_{rx}}{k_0}\frac{\partial r_{pp}}{\partial \theta}\right)|{H}({k}_{r})\rangle+\frac{k_{ry}\cot\theta(r_{sp}-r_{ps})}{k_{0}}|{H}({k}_{r})\rangle\nonumber\\&&-\frac{k_{ry}\cot\theta(r_{pp}+r_{ss})}{k_{0}}|{V}({k}_{r})\rangle
 + \left(r_{sp}-\frac{k_{rx}}{k_0}\frac{\partial r_{sp}}{\partial \theta}\right)|{V}({k}_{r})\rangle\nonumber.\\\label{HKI}
\end{eqnarray}
Here, $k_0=\omega/c$ is the wavevector in vacuum and $\theta$ is the angle of incidence.

From the relations of $|H\rangle=\frac{1}{\sqrt{2}}(|{+}\rangle+|{-}\rangle)$ and $|V\rangle=\frac{1}{\sqrt{2}}i(|{-}\rangle-|{+}\rangle)$, we next analyze Eq.~(\ref{HKI}) in spin basis to reveal the splitting of spin components. $|{+}\rangle$ and $ |{-}\rangle$ represent the left- and right-circular polarization components, respectively. Supposing we have $r_{ps}=r_{sp}$, the total momentum wavefunction in the spin basis is
\begin{eqnarray}
  |{\psi_{\pm}}\rangle&=&\frac{1}{\sqrt{2}}\bigg[r_{pp}-\frac{k_{rx}}{k_0}\frac{\partial r_{pp}}{\partial \theta}\mp i\left(r_{sp}-\frac{k_{rx}}{k_0}\frac{\partial r_{sp}}{\partial \theta}\right)\nonumber\\&&\pm i\frac{k_{ry}(r_{pp}+r_{ss})\cot\theta}{k_0}\bigg] |\pm\rangle|\Phi\rangle\label{WFH}.
  \end{eqnarray}
Considering the incident beam with a Gaussian distribution, $|\Phi\rangle$ is given by
\begin{equation}
|\Phi\rangle=\frac{w_{0}}{\sqrt{2\pi}}\exp\left[-\frac{w^{2}_{0}(k_{ix}^{2}+k_{iy}^{2})}{4}\right]\label{GaussianWF},
\end{equation}
where $w_{0}$ is the width of wavefunction. Taking into account the paraxial approximation, the wavefunction expression then can be simplified as
 \begin{eqnarray}
  |{\psi}_{\pm}\rangle&=&\frac{r_{pp}\mp ir_{sp}}{\sqrt{2}}(1\pm ik_{rx}\delta_{\pm}^{x}\pm ik_{ry}\delta_{\pm}^{y}) |\pm\rangle|\Phi\rangle\nonumber\\
  &&\approx\frac{r_{pp}\mp ir_{sp}}{\sqrt{2}}\left( e^{\pm ik_{rx}\delta_{\pm}^{x}}e^{\pm ik_{ry}\delta_{\pm}^{y}}\right) |\pm\rangle|\Phi\rangle\label{W},
  \end{eqnarray}
 in which $\delta_{\pm}^{x}=(\partial r_{sp}/\partial \theta\pm i \partial r_{pp}/\partial \theta)/[k_{0}(r_{pp}\mp i r_{sp})]$ and $\delta_{\pm}^{y}=[(r_{pp}+r_{ss})\cot\theta]/[k_0(r_{pp}\mp ir_{sp})]$.  A straightforward calculation based on
\begin{equation}
\langle{\Delta_{\pm}^{x, y}}\rangle=\frac{\langle\psi_{\pm}|\partial_{k_{rx, ry}}|\psi_{\pm}\rangle}{\langle\psi_{\pm}|\psi_{\pm}\rangle},\quad \langle{\Theta_{\pm}^{x, y}}\rangle=\frac{1}{k_0}\frac{\langle\psi_{\pm}|{k_{rx, ry}}|\psi_{\pm}\rangle}{\langle\psi_{\pm}|\psi_{\pm}\rangle}\label{PYHV}.
\end{equation}
can yields the in-plane spatial and angular shifts as
 \begin{eqnarray}
  \langle \Delta_{\pm}^{x}\rangle&=&\mp \mathrm{Re}\left[\frac{\partial r_{sp}/\partial \theta\pm i \partial r_{pp}/\partial \theta}{(r_{pp}\mp ir_{sp})k_0}\right]\label{C},
  \end{eqnarray}
   \begin{eqnarray}
  \langle \Theta_{\pm}^{x}\rangle&=&\mp \frac{1}{z_R} \mathrm{Im}\left[\frac{\partial r_{sp}/\partial \theta\pm i \partial r_{pp}/\partial \theta}{(r_{pp}\mp ir_{sp})k_0}\right]\label{D},
  \end{eqnarray}
where the $z_R$ is the Rayleigh length. A similar result can be obtained for the transverse direction
 \begin{eqnarray}
  \langle \Delta_{\pm}^{y}\rangle&=&\mp \mathrm{Re}\left[\frac{(r_{pp}+r_{ss})\cot\theta}{(r_{pp}\mp ir_{sp})k_0}\right]\label{A},
  \end{eqnarray}
   \begin{eqnarray}
  \langle \Theta_{\pm}^{y}\rangle&=&\mp \frac{1}{z_R} \mathrm{Im}\left[\frac{(r_{pp}+r_{ss})\cot\theta}{(r_{pp}\mp ir_{sp})k_0}\right]\label{B},
  \end{eqnarray}

The results in Eqs.~(\ref{C}) - (\ref{B}) is the simplest form to describe the behavior of spin-orbit interaction of light. The real and imaginary parts of $\delta_{\pm}^{x, y}$ correspond to the spatial and angular shifts of the two spin components~\cite{Cai2017}. Thus, the reflection coefficients play a crucial role in demonstrating the spin-orbit interaction of light. Moreover, the reflection coefficients is related to the preselected state in weak measurement schema, which will be clear in Sec. III.

 \begin{figure}
\includegraphics[width=8cm]{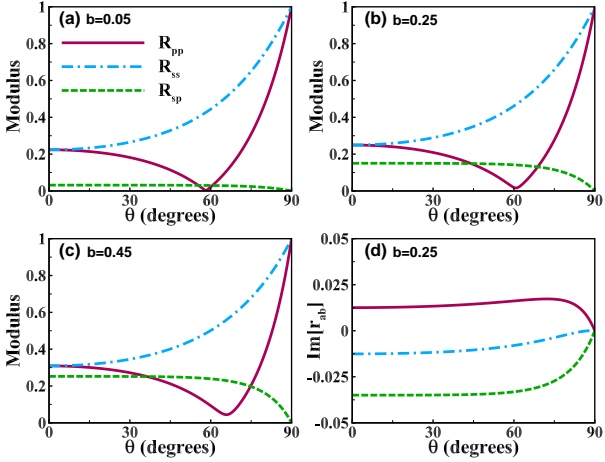}
\caption{\label{Fig3} Modulus of reflection coefficients of the WSM-substrate system as a function of incident angle $\theta$ for (a) $b=0.05$, (b) $b=0.25$, and (c) $b=0.45$.  (d) Imaginary part of the coefficients with $b=0.25$. Parameters for WSM are chosen as $d$=10nm, $v_{F}$=$10^{6}$m/s, and $a$=3.44{\AA}, and the refractive index of substrate is $n$=1.5.}
\end{figure}

We next detailedly discuss the reflection coefficients together with the Weyl nodes to give a insight into the interaction of light. To obtain the Fresnel reflection coefficients in WSM-substrate system, the boundary conditions for electromagnetic field and the Ohm's law should be taken into account~\cite{Tse2011,Kamp2015,Merano2016}. Assuming the electric (magnetic) fields in air and substrate are respectively represented by $\textbf{E}_{1}$ and $\textbf{E}_{2}$ ($\textbf{H}_{1}$ and $\textbf{H}_{2}$), the boundary conditions are $\hat{n}\times(\textbf{E}_{2}-\textbf{E}_{1})=0$, $\hat{n}\times(\textbf{H}_{2}-\textbf{H}_{1})=\textbf{J}_{s}$. $\hat{n}=-\hat{z}$ is the unit vector normal to the WSM-substrate interface, and $\textbf{J}_{s}=\sigma_{\imath\jmath}\textbf{E}$ is the surface current density. $\sigma_{\imath\jmath}$ denotes the surface conductivity tensor in WSM with $\imath,\jmath=x,y$. Solving the boundary condition expressions, we get the coefficients for arbitrary incident angles as
  \begin{equation}
  r_{pp}=\frac{\alpha^T_+\alpha_-^L+\beta}{\alpha_+^T\alpha_+^L+\beta}\label{RPP},
  \end{equation}
  \begin{equation}
  r_{ss}=-\frac{\alpha^T_-\alpha_+^L+\beta}{\alpha^T_+\alpha_+^L+\beta}\label{RSS},
  \end{equation}
   \begin{equation}
   r_{ps}=r_{sp}=-2\sqrt{\frac{\mu_0}{\varepsilon_0}}\frac{k_{iz}k_{tz}\sigma_{xy}}{\alpha^T_+\alpha_+^L+\beta}\label{RPS},
   \end{equation}
  where $\alpha^L_\pm=(k_{iz}\varepsilon\pm k_{tz}\varepsilon_0+k_{iz}k_{tz}\sigma_{xx}/\omega)/\varepsilon_0$,
  $\alpha^T_\pm=k_{tz}\pm k_{iz}+\omega\mu_0\sigma_{yy}$, and $\beta=\mu_0k_{iz}k_{tz}\sigma^2_{xy}/\varepsilon_0$. $k_{iz}=k_0\cos\theta$ and $k_{tz}=nk_0 \cos\theta_t$.
  $\theta_t$ is the refraction angle; $n$ is the refractive index of the substrate; $\varepsilon_0$ , $\mu_0$  are
  permittivity and permeability in vacuum; $\varepsilon$ is the permittivity of substrate;
  $\sigma_{xx,yy}$and $\sigma_{xy, yx}$
  are the longitudinal and Hall conductivities, respectively.

In our case, the WSM film is ultra-thin ($a\ll d\ll\lambda$). Two Weyl nodes are separated by a wave vector $\pm\textbf{b}=\pm(0, 0, b)$ in the Brillouin zone. $b$ is in units of 2$\pi/a$ throughout this paper with $a$ representing the lattice spacing. For the WSM thin film, the corresponding optical conductivity is $\sigma_{\imath\jmath}=d\sigma^{B}_{\imath\jmath}$. $\sigma^{B}_{\imath\jmath}$ is the conductivity of the bulk WSM obtained from the Kubo formalism~\cite{Kargarian2015,Ahn2017}. The real and imaginary parts of the optical conductivity $\sigma^{B}_{\imath\jmath}$ are given by
\begin{eqnarray}
\mathrm{Re}[\sigma_{xx}^{B}]&=&\frac{e^{2}|\omega|}{12\pi\hbar v_{F}}\label{lb},\\
\mathrm{Im}[\sigma_{xx}^{B}]&=&-\frac{e^{2}\omega}{12\pi^{2} \hbar v_{F}}\ln\left|\frac{\omega^{2}-\omega^{2}_{c}}{\omega^{2}}\right|,\\
\mathrm{Re}[\sigma_{xy}^{B}]&=&be^{2}\left[\frac{1}{2\pi^2 \hbar}+\frac{\omega^{2}}{24\pi^2 \hbar v^{2}_{F}(k_{c}^{2}-b^{2})}\right],\\
\mathrm{Im}[\sigma_{xy}^{B}]&=&-\left[\frac{e^{2}b}{2\pi^{3}\hbar}+\frac{e^{2}\omega^{2}b}{24\pi^{3}\hbar v_{F}^{2}(k_c^2-b^2)}\right]\ln\left|\frac{\omega_c-\omega}{\omega_c+\omega}\right|\nonumber\\&&-\frac{e^{2}\omega\omega_c b}{12\pi^{3}\hbar v_{F}^{2}(k_c^2-b^2)}\label{lk}.
\end{eqnarray}
 For detailed calculation to the optical conductivity, one can see Appendix. The conductivity of WSM shows a characteristic frequency dependence. Note that only the positive frequencies of $\sigma_{xx}^{B}$ is discussed in Appendix, and Eqs.~(\ref{lb}) - (\ref{lk}) are hold at the low frequency limit $\omega<\omega_{c}$~\cite{Hosur2012,Ashby2013}. The corresponding theoretical predictions for the optical conductivity have been experimentally verified~\cite{Xu2016}. For other metals such as topological insulators, the dynamic conductivity arises as a function of temperature and photon energy in the surface states~\cite{Li2015,Li2015I}.  We see that the Hall conductivity brings about the influence of the Weyl nodes.  If $b=0$, indicating the annihilation of Weyl nodes in reflection coefficients, the conductivity $\sigma_{xy}$ vanishes. And the Fresnel reflection coefficients reduce to a general case~\cite{Merano2016}.

 Due to the complex optical conductivities, the reflection coefficients associated with the location of Weyl nodes are complex numbers. We write the coefficients as $r_{ab}=R_{ab}e^{i\varphi_{ab}}$ with $R_{ab}$ and $\varphi_{ab}$ labeling modulus and phase, respectively. In Fig.~\ref{Fig3}, the modulus for three different distances of the Weyl nodes are plotted as a function of incident angle.  For small separation of the Weyl nodes ($b=0.05$), the behaviors of the reflection coefficients are nearly the same as the case without WSM film. With the existence of WSM, the angle of $r_{pp}=0$ vanishes. Such a angle in the case with zero crossing reflection coefficients is known as the Brewster angle. Near this incident angle, the action of spin-orbit interaction may become peculiar, such as resulting in a very large spin-dependent splitting. As the separation increases, $r_{sp}$ and $r_{pp}$ become large gradually. But the influence of the Weyl nodes to $r_{ss}$ is not obvious. To show the contribution of the imaginary parts to the reflection coefficients, Fig.~\ref{Fig3}(d) is provided for $b=0.4$. The imaginary parts of the reflection coefficients is not small enough to be neglected. We point out that for other $b$ there also exits non-negligible imaginary part in the optical conductivity.

\begin{figure}
\includegraphics[width=8cm]{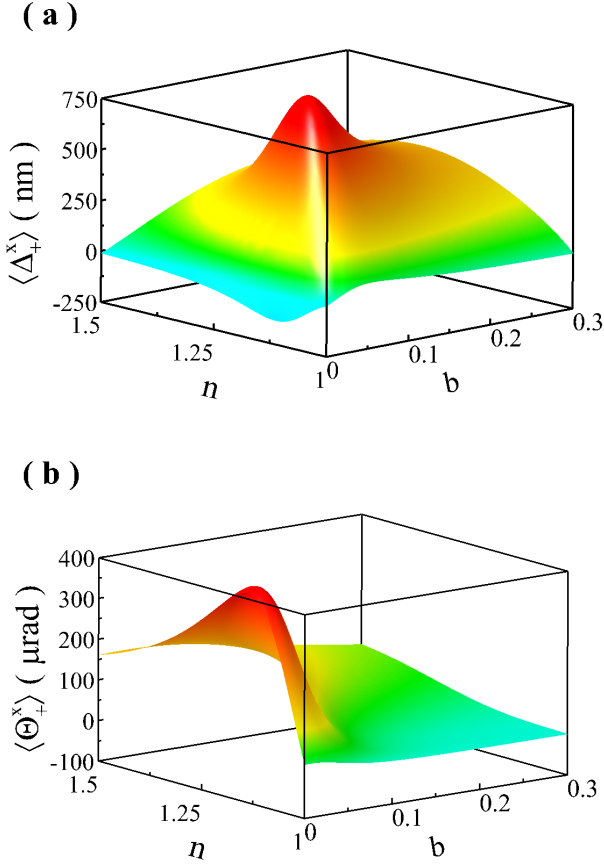}
\caption{\label{Fig4} In-plane (a) spatial and (b) angular shifts as a function of refractive index of substrate $n$ and parameter $b$ (in units of 2$\pi/a$) associated with the location of Weyl node. We assume an incident beam with $|H\rangle$ polarization, $w_{0}$=27$\mu$m, and $\lambda$=633nm. The angle of incidence of the light beam is set as 50 degrees. Parameters for WSM are the same as in Fig.~\ref{Fig3}.}
\end{figure}

\begin{figure}
\includegraphics[width=8cm]{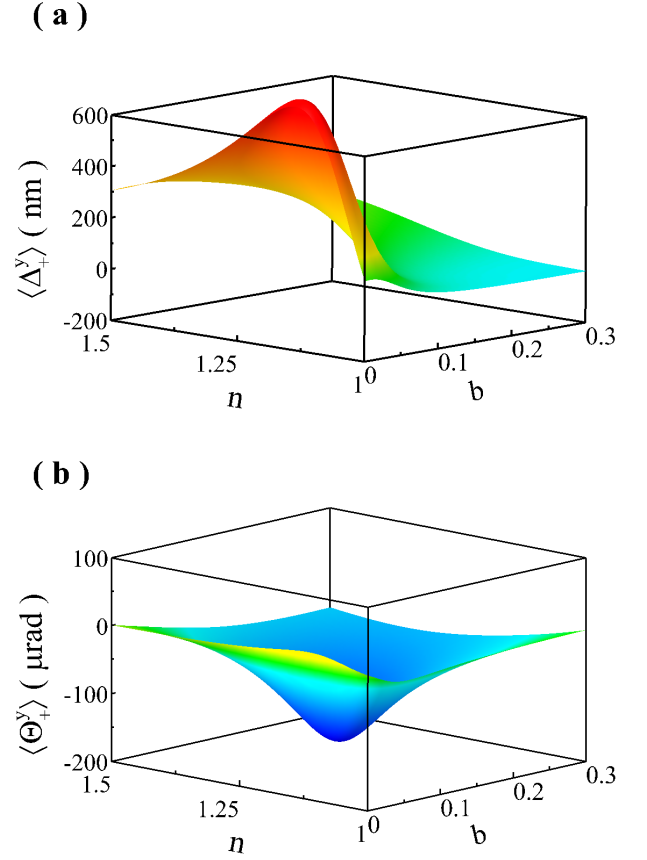}
\caption{\label{Fig5} Transverse (a) spatial and (b) angular shifts as a function of refractive index of substrate $n$ and parameter $b$. The parameters are the same as in Fig.~\ref{Fig4}.}
\end{figure}

 The spatial and angular shifts are related to the real and imaginary parts of $\delta_{\pm}^{x, y}$. Based on the result of reflection coefficients being complex numbers, one shift in Eqs.~(\ref{C}) - (\ref{B}) contains both spin-independent and -dependent components. To show how the refractive index of substrate impact on the spin-orbit interaction of light in a WSM-substrate system, we first discuss the shifts as a function of refractive index of substrate $n$ and parameter $b$. We only plot the shifts of left handed circular component. In Fig.~\ref{Fig4}, our result shows that the substrate can effectively influence the shifts. Both in-plane spatial and angular shifts exhibit a peak value at $n\approx1.13$. Such a condition may be helpful for the investigation of Weyl nodes. For $n>1.5$, the shifts become very small. At about $b=0.04$, there exits a large peak about 800 nm for spatial shift. And the angular shift becomes maximal with $b=0$.

\begin{figure}
\includegraphics[width=8cm]{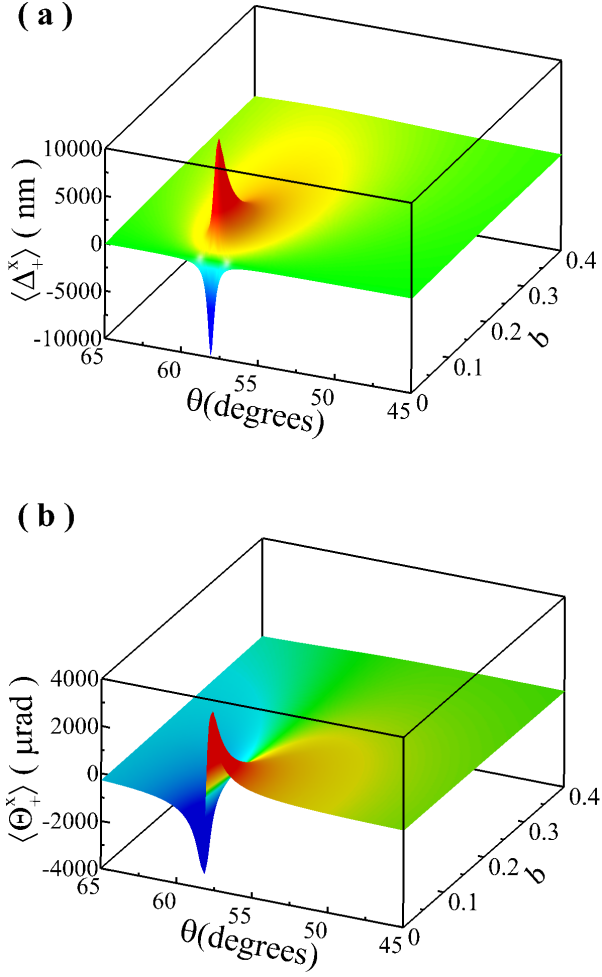}
\caption{\label{Fig6} In-plane (a) spatial and (b) angular shifts as a function of incident angle $\theta$ and parameter $b$ when the refractive index of substrate is set as $n$=1.5.}
\end{figure}

\begin{figure}
\includegraphics[width=8cm]{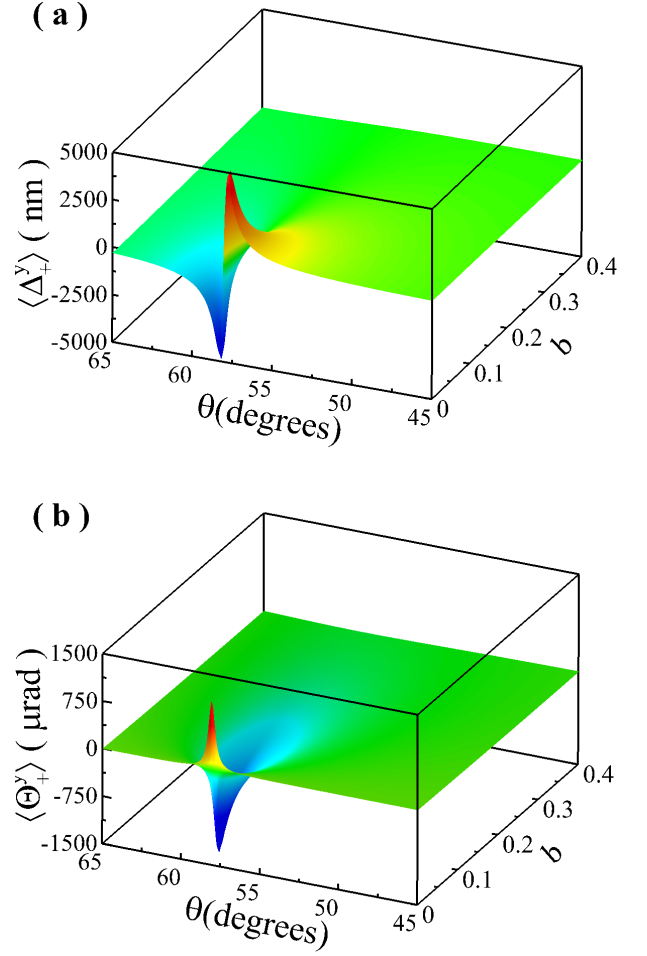}
\caption{\label{Fig7} Transverse (a) spatial and (b) angular shifts as a function of incident angle $\theta$ and parameter $b$. The parameters are the same as in Fig.~\ref{Fig6}.}
\end{figure}

For the case of transverse shifts, it also exists the optimal substrate refractive index $n\approx1.13$ to obtain strong spin-orbit interaction of light. Without the Weyl nodes, the transverse spatial shift can be very large. In fact, a system without WSM, such as the air-glass interface, can lead to the transverse spatial shift as well~\cite{Hosten2008,Qin2009,Luo2011,Chen2015}. The WSM coating only affects the magnitude of the transverse splitting with different separation of Weyl nodes. From Eq.~(\ref{B}), the WSM coating is also responsible for the existence of transverse angular shift due to the nonzero cross-polarization coefficient and the complex $r_{pp}$ and $r_{ss}$. The angular shift at $b=0$ with $r_{sp}=0$ in Fig.~\ref{Fig5}(b) is nearly zero. The maximum value of transverse angular shift appears at $b\approx0.04$.

We next discuss the shifts related to incident angle $\theta$ and parameter $b$. Incident angle is an important factor to influence the coupling strength of the spin-orbit interaction of light. The result in Eqs.~(\ref{C}) - (\ref{B}) is the simplest form to describe the behavior of spin-orbit interaction of light. If the separation of the Weyl nodes becomes zero, the terms containing $r_{ps}$ vanish, and the expressions are invalid near the Brewster angle. Under such a situation, the small variations of the Fresnel coefficients should be taken into account. By substituting Eqs.~(\ref{WFH}) and (\ref{GaussianWF}) into Eq.~(\ref{PYHV}), precise shifts at initial position ($z_r=0$) can be obtained. The precise result only modifies the shift in the limit that the Weyl points disappear ($b=0$). For $b>0$, it is almost the same as the one given by the approximate formula.

From Eq.~(\ref{C}), the in-plane spatial shift originates from the nonzero coefficient $r_{sp}$ and the complex $r_{pp}$. $r_{sp}$ is a quantity related to the separation of the Weyl point. A giant in-plane shift about $12\mu$m at $b\approx0.02$ can be realized in the vicinity of the Brewster angle, as shown in Fig.~\ref{Fig6}(a). This is much larger than the case of transverse shift in Fig.~\ref{Fig7}(a), in which the shift is about $3500$nm at 57.5degrees, and its maximal value reaches only about $5000$nm. The sensitivity of the shifts to the parameters of WSM near the Brewster angle can improve precision during the weak measurement progress~\cite{Zhou2012I,Zhou2013}. The sharp peak for the spatial shift occurs at $\theta\approx58$degrees due to the term $(r_{pp}^{2}+r_{sp}^{2})\thickapprox0$ in Eqs.~(\ref{C}) and ~(\ref{A}). For both in-plane and transverse angular shifts, $b=0$ can also lead to a peak value near the Brewster angle. But there is another peak of the transverse angular shift appears at $b\approx0.02$, like the case of the in-plane spatial shift.

\section{Quantum weak value amplification}\label{SecIII}

In this section, we introduce the quantum weak measurements to observe this tiny effect. In a weak measurement scheme, only the relative component can be amplified by weak value. Thus, only the spin-dependent component in our case can be filtrated for detection. Theoretically, both the real and imaginary parts of weak value can enhance the tiny observable. For example, a spatial shift amplified by the real and imaginary parts of the weak value corresponds to the position and momentum shifts, respectively~\cite{Hosten2008,Aiello2008,Chen2017}. The weak value is naturally determined by preselected and postselected states. In our case, the preselected state $|{i}\rangle$ is the polarization after interacting with the WSM-substrate interface. We obtain it as $|{i}\rangle=F|{H}\rangle$ with the incident state $|{H}\rangle$. $F$ is the reflection matrix of WSM-substrate system
\begin{eqnarray}
\left[
\begin{array}{cc}
r_{pp} &r_{ps} \\
r_{sp} &
r_{ss}
\end{array}\right].
\end{eqnarray}
The preselected state in the spin basis is
\begin{equation}
|{i}\rangle=\cos\upsilon|+\rangle+e^{i\gamma}\sin\upsilon|-\rangle,
\end{equation}
where $\upsilon=\arccos\left(|r_{pp}-i r_{sp}|/\sqrt{|r_{pp}-i r_{sp}|^2+|r_{pp}+i r_{sp}|^2}\right)$  and $\gamma=\arg(r_{pp}-i r_{sp})-\arg(r_{pp}+i r_{sp})$. In order to reach a large weak value, the postselected state needs to be nearly orthogonal to $|{i}\rangle$. Suppose it is chosen as
\begin{equation}
|{f}\rangle=\sin\upsilon|+\rangle-e^{i(\gamma-2\phi)}\cos\upsilon|-\rangle\label{MIDS},
\end{equation}
in which the small deviation angle $\phi$ is also called as postselected angles. From above preselected and postselected states, the weak value $\sigma_w$ is calculated as~\cite{Jordan2014,Chen2017}
\begin{equation}
\sigma_{w}=\frac{\langle {f}|\sigma_3|{i}\rangle}{\langle
{f}|{i}\rangle}=i\cot\phi,\label{AWI}
\end{equation}
where $\sigma_3$ is the Pauli operator. This purely imaginary weak value only effectively amplifies the spatial shift due to the free evolution of the wave function in the momentum space. For the angular shift, the amplified factor is small. The total amplification result is obtained as
 \begin{equation}
 \langle{x, y}\rangle=\left(\frac{z_r}{z_R} \frac{\mid\langle\Delta_{+}^{x, y}\rangle-\langle\Delta_{-}^{x, y}\rangle\mid}{2}-\frac{z_R\mid\langle\Theta_{+}^{x, y}\rangle-\langle\Theta_{-}^{x, y}\rangle\mid}{2}\right)\cot\phi\label{AS}.
 \end{equation}
 The results in Eq.~(\ref{AS}) is propagation-dependent and the shifts can be effectively amplified in far field. On the other side, to achieve a large $\sigma_{w}$, one can make $\phi\rightarrow0$ as much as possible. But in fact the shift has a maximum value when the postselected angle is close to zero. Under this situation, a modified weak measurements is required~\cite{Chen2015}.

  \begin{figure}
\includegraphics[width=8cm]{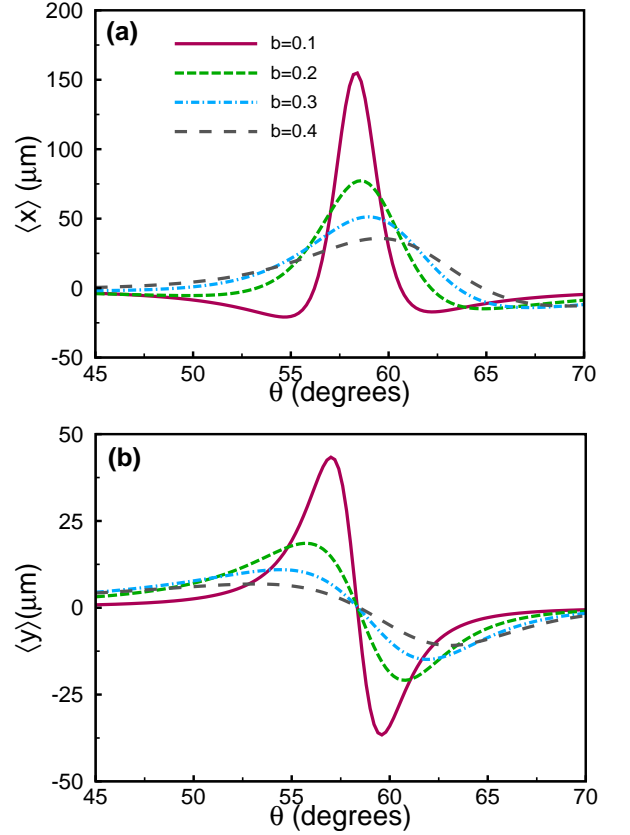}
\caption{\label{Fig8} (a) In-plane and (b) transverse shifts amplified by purely imaginary weak value are shown as
functions of incident angle $\theta$. The postselected angle $\phi$ is chosen as 0.2 degrees, and propagation distance $z_r$ is 1000mm.}
\end{figure}

We plot the amplified shifts as functions of incident angle $\theta$ in Fig.~\ref{Fig8} by setting the postselected angle $\phi$ as 0.2 degrees. A large amplified factor ${z_r}\cot\phi/{z_R}\thickapprox79200$ can be obtained for spatial shifts.  Assuming the weak value is real, the amplified factor $\cot\phi$ is only about 286 due to no propagation enlargement. Therefore, the imaginary weak value here is a good choice to detect the spatial displacement. We discuss four different separations of Weyl nodes, $b=0.1$, $b=0.2$, $b=0.3$, and $b=0.4$. The result shows that the in-plane and transverse shifts amplified by this factor can be reach dozens of micrometers, which is detectable in the experiment. The curves become steeper near the Brewster angle, especially for the case with small $b$. For $b=0.1$, a peak of in-plane shift about 150$\mu$m is achieved. The sensitivity of the amplified shift with increasing parameter $b$ makes it feasible to detect the Weyl nodes in the experiment. Near the Brewster angle, the difference between two curves can reach about 120$\mu$m. The change of the shift with $b$ become more sensitive when the Weyl point separation is small. This situation may be helpful to determine $b$, or even other parameters of WSM. Recently optical experiments on WSMs are still very limited~\cite{Xu2016,Ma2017}. The wave length we consider here is at the visible range,which is accessible in an experiment.

\section{Conclusions}

In conclusion, we have theoretically discuss the spin-orbit interaction of light reflected on the WSM-substrate interface. We predict both in-plane and transverse shifts with the presence of WSM. The WSM we consider contains a pair of Weyl nodes, and the light incidents on the surface of WSM that does not support Fermi-arc electronic states. The analysis shows that the spin-dependent in-plane spatial and transverse angular shifts originate from the existence of Weyl points. Introducing a purely imaginary weak value, the spatial shifts can be effectively enlarged by a factor of  ${z_r}|\sigma_{w}|/{z_R}\approx8\times10^{4}$, which is 276 times larger than the one with real weak value. Due to the sensitivity to the wave vector $\pm\textbf{b}$, measuring beam shift could become an alternative way to determine the distance of Weyl nodes in momentum space. Our results may open up a new experimental possibility for the investigations of optical responses into WSM.

\section*{ACKNOWLEDGMENTS}
The authors are sincerely grateful to Dr. Qinjun Chen for many fruitful discussions. This research was supported by the National Natural Science Foundation of China (Grant No. 11474089); Hunan Provincial Innovation Foundation for Postgraduate (Grant No. CX2016B099).

 \begin{appendix}
\section{Calculation of the optical conductivity}
We obtain the optical conductivities $\sigma_{xx}^{B}$ and $\sigma_{xy}^{B}$ of the WSM from the Kubo formula in the noninteracting limit
\begin{eqnarray}
  \sigma_{\imath\jmath}^{B}(\omega)&=&-\frac{ie^{2}}{\hbar}\sum\limits_{s,s^{'}}\int\frac{d^{3}k}{(2\pi)^{3}}\frac{f_{s,k}-f_{s^{'},k}}{E_{s,k}-E_{s^{'},k}}\nonumber\\
  &&\times\frac{M^{ss^{'}}_{\imath}M^{s^{'}s}_{\jmath}}{\omega\hbar+E_{s,k}-E_{s^{'},k}+i0^{+}}\label{Ku},
  \end{eqnarray}
  where $f_{s,k}=1/[1+e^{(E_{s,k}-\mu)/k_B T}]$ is the Fermi distribution function with $\mu$ being the chemical potential, and  $M^{ss^{'}}_{\imath}=\langle s|\hbar\hat{v_{\imath}}|s^{'}\rangle$. $\hat{v_{\imath}}$ is the velocity operator that can be obtained from the Hamiltonian with the relation of $\hat{v_{\imath}}=\frac{1}{\hbar}\frac{\partial\hat{H}}{\partial k_\imath}$.

  Considering the case with only two Weyl nodes located at $\pm\textbf{b}=b\hat{z}$, the corresponding low-energy Hamiltonian is given by~\cite{Ahn2017}
  \begin{equation}
H=E_{0}\left(\frac{k_{-}\sigma_{+}}{k_{m}}+\frac{k_{+}\sigma_{-}}{k_{m}}\right)+\hbar v_{F} q_{z}\sigma_{z},\label{Ha}
\end{equation}
where $k_{\pm}=k_x\pm ik_y$, $\sigma_{\pm}=\frac{1}{2}(\sigma_x\pm i\sigma_y)$ with $\sigma$ representing the Pauli matrices, and $q_z=k_z\mp b$ is the effective wave vector along the $k_z$ direction. $E_{0}$ and $k_{m}$ are material-dependent parameters related to energy and momentum, respectively. For brevity, we set them as $E_{0}=k_{m}=1$ throughout this section.

We calculate the optical conductivities in the clean limit at zero temperature. For the longitudinal conductivity, the intraband contribution of the optical conductivity in the undoped case ($\mu=0$) can be neglected. At positive frequencies, the real part of the conductivity from interband contribution can be obtained by
\begin{eqnarray}
\mathrm{Re}[\sigma_{xx}^{B}(\omega)]&=&-\frac{\pi e^{2}}{\hbar}\int\frac{d^{3}k}{(2\pi)^{3}}\frac{f_{-,k}-f_{+,k}}{E_{-,k}-E_{+,k}}\nonumber\\&&\times|M_{x}^{-+}|^{2}\delta(\omega\hbar+E_{-,k}-E_{+,k})\label{xx}
\end{eqnarray}
with $M^{ss^{'}}_{\imath}M^{s^{'}s}_{\imath}=|M^{ss^{'}}_{\imath}|^{2}$.
A straightforward calculation from Eq.~(\ref{xx}) yields
\begin{equation}
\mathrm{Re}[\sigma_{xx}^{B}(\omega)]=\frac{e^{2}}{12\pi\hbar v_{F}}\omega.\label{xx1}
\end{equation}
For the case of both positive and negative frequencies, the form of $\sigma_{xx}^{B}(\omega)$ is given by Eq.~(\ref{lb})

Using Eq.~(\ref{Ku}), we can also obtain the real part of the Hall or transverse optical conductivity for $\mu=0$ as
\begin{eqnarray}
&\mathrm{Re}&[\sigma_{xy}^{B}(\omega)]=-\frac{ie^{2}}{\hbar}\int\frac{d^{3}k}{(2\pi)^{3}}\frac{f_{+,k}-f_{-,k}}{E_{+,k}-E_{-,k}}\nonumber\\&&\times\left[\frac{M_{x}^{+-}M_{y}^{-+}}{\omega\hbar+E_{+,k}-E_{-,k}}+\frac{M_{x}^{-+}M_{y}^{+-}}{\omega\hbar+E_{-,k}-E_{+,k}}\right].\label{xy}
\end{eqnarray}
In the limit of low frequencies, the conductivity with the momentum cut-off $k_c$ along $k_z$ axis is calculated as
\begin{equation}
\mathrm{Re}[\sigma_{xy}^{B}(\omega)]=\frac{e^{2}b}{2\pi^2 \hbar}+\frac{e^{2}b}{24\pi^2 \hbar v^{2}_{F}(k_{c}^{2}-b^{2})}\omega^{2}.\label{xy1}
\end{equation}
In the following, we obtain the imaginary part of the conductivity by introducing the Kramers--Kronig relation~\cite{Landau}. The transformation expression is given by
\begin{equation}
\mathrm{Im}[\sigma_{\imath\jmath}^{B}(\omega)]=-\frac{2\omega}{\pi}\int^{\infty}_{0}\frac{\mathrm{Re}[\sigma_{\imath\jmath}^{B}(\omega^{'})]}{\omega^{'2}-\omega^{2}}d \omega^{'}.\label{KK}
\end{equation}
This formula is valid in the condition of $\sigma_{\imath\jmath}(-\omega)=\sigma^{*}_{\imath\jmath}(\omega)$. Substituting Eq.~(\ref{xx1}) into Eq.~(\ref{KK}), the imaginary part of the optical conductivity $\sigma_{xx}^{B}$ with a cut-off $\omega_{c}$ is given by
 \begin{equation}
\mathrm{Im}[\sigma_{xx}^{B}(\omega)]=-\frac{e^{2}\omega}{12\pi^{2}\hbar v_{F}}\ln\left|\frac{\omega^{2}-\omega^{2}_c}{\omega^{2}}\right|.\label{xx2}
\end{equation}
Similarly, we can obtain the imaginary part of $\sigma_{xy}^{B}$ from its real part as
 \begin{eqnarray}
\mathrm{Im}[\sigma_{xy}^{B}(\omega)]&=&-\left[\frac{e^{2}b}{2\pi^{3}\hbar}+\frac{e^{2}\omega^{2}b}{24\pi^{3}\hbar v_{F}^{2}(k_c^2-b^2)}\right]\ln\left|\frac{\omega_c-\omega}{\omega_c+\omega}\right|\nonumber\\&&-\frac{e^{2}\omega\omega_c b}{12\pi^{3}\hbar v_{F}^{2}(k_c^2-b^2)}.\label{xy2}
\end{eqnarray}
\end{appendix}

\end{document}